\documentclass[aps,prl,superscriptaddress,showpacs,preprintnumbers,twocolumn]{revtex4}

\usepackage{graphicx}
\usepackage{dcolumn}
\usepackage{bm}
\usepackage{pstricks}

\def\va1{\vec{a}_{1}}

\def\vb1{\vec{b}_{1}}

\def\vd1{\vec{\delta}_{1}}

\newcommand{\ba}{\begin{eqnarray}}
\newcommand{\ea}{\end{eqnarray}}
\def\be{\begin{equation}}
\def\ee{\end{equation}}

\def\vk{\mathbf{k}}
\def\vsigma{\textbf{k}}

\newcommand{\step}[2]{\vskip.02cm{\noindent{{\bf Step #1: }}{#2}}\vskip.7cm}

\begin{document}

\title{Generalized Pomeranchuk instabilities in graphene}
\author{C.A. Lamas}
\affiliation{Departamento de F\'isica, Universidad Nacional de La
Plata, Casilla de Correos 67, 1900 La Plata, Argentina}
\author{D.C. Cabra}
\affiliation{Departamento de F\'isica, Universidad Nacional
de La Plata, Casilla de Correos 67, 1900 La Plata, Argentina}
\affiliation{Laboratoire de Physique Th\'{e}orique, Universit\'{e}
Louis Pasteur, 3 Rue de l'Universit\'{e}, 67084 Strasbourg, C\'edex,
France.}
 \affiliation{Facultad de Ingenier\'\i a, Universidad
Nacional de Lomas de Zamora, Cno.\ de Cintura y Juan XXIII, (1832)
Lomas de Zamora, Argentina.}
\author{N. Grandi}
\affiliation{Departamento de F\'isica, Universidad Nacional de La
Plata, Casilla de Correos 67, 1900 La Plata, Argentina}
%

\begin{abstract}
We study the presence of Pomeranchuk instabilities induced by
interactions on a Fermi liquid description of a graphene layer. Using
a recently developed generalization of Pomeranchuk method we present
a phase diagram in the space of fillings versus on-site and nearest
neighbors interactions. Interestingly, we find that for both
interactions being repulsive an instability region exists near the
Van Hove filling, in agreement with earlier theoretical work. In
contrast, near half filling, the Fermi liquid behavior appears to be
stable, in agreement with theoretical results and experimental
findings using ARPES. The method allows for a description of the complete phase diagram for
arbitrary filling.

\end{abstract}

\maketitle

\section{Introduction}
\label{sec:derivation}

Correlated electron systems in two dimensions have attracted a lot
of attention in the last years, especially due to an important number of
experiments that provide undisputable evidence of the existence of
new exotic phases of matter.

One such example corresponds to nematic
and stripe (smectic) phases in high $T_c$ superconductors in the
underdoped region and fractional quantum Hall effect systems at high
magnetic fields \cite{Fradkin_nematic}. A nematic phase is
characterized by orientational but not positional order and it has
been proposed to explain the observed transport anisotropies. One
important point about these phases is that they arise spontaneously,
decreasing the rotational symmetry without a lowering of the lattice
symmetry. Another more recent case is given by strontium ruthenate
Sr$_3$Ru$_2$O$_7$, which is well modeled as a bilayer system and
shows a large magnetoresistive anisotropy. This observation has
been argued to be consistent with an electronic nematic fluid phase.
Experimentally, two consecutive metamagnetic transitions have been
observed and the region in between has been proposed to be
a consequence of a Pomeranchuk instability, due to a nematic
deformation of the Fermi surface, in very close analogy to what
happens in fractional quantum Hall gallium arsenide systems \cite{Borzi et al}.
Yet another interesting material is the heavy fermion compound
URu$_2$Si$_2$ in which a hidden order phase arises through a second
order transition at around $17.5^0 $K. The order parameter of this
new phase has remained elusive to theorists up to date. Different
types of order have been proposed, but the situation is still under
debate \cite{variousorders}. In recent work, Varma and Zhu
\cite{VarmaZhu} have proposed that this phase transition could
correspond to a Pomeranchuk instability inducing a deformation in the antisymmetric spin
channel, stabilized by a phase characterized by a helicity order
parameter.

The experimental findings mentioned above triggered different
theoretical studies on low dimensional correlated systems to search
for such exotic phases \cite{Fradkin_nematic}. Special attention has
been paid to the possibility of Pomeranchuk instabilities \cite{Pomeranchuk} giving
rise to such novel phases \cite{Metzner1,Metzner2,Varma,Yamasenew,Quintanilla1,Quintanilla2,Wu,Nilson,polacos,Belen1,Gonzalez}. In a previous paper
\cite{LCG}, motivated by these investigations, we developed a
generalization of Pomeranchuk's method to search for instabilities
of a Fermi liquid. The method we presented is
applicable to any two dimensional lattice model with an arbitrary
shape of the Fermi surface (FS) at zero temperature. The main results of our
previous paper were summarized in the form of a recipe whose steps
we give below for completeness. Our method is particularly well suited
to analyze systems with weak interactions and then graphene appears
as a perfect arena to test our techniques, since electron
interactions are argued to be small due to strong screening.

The recent isolation of graphene \cite{Novoselov}, the first purely
two-dimensional material, which is made out of carbon atoms arranged
in a hexagonal structure, led to an enormous interest and a large amount of activity
in studying its properties. The low doping region, near to half
filling, became the subject of attention due to the peculiar
behavior described by chiral massless charge carriers. Several Van
Hove singularities are present at energies of the order of the
hopping parameter $E\sim 2.7$ eV and these singularities are
expected to have an important role in the properties of the system.
Although in first approximation graphene layers are well modeled
by free fermions hopping on a hexagonal lattice, there have been a
number of works in the literature where the effects of
electron-electron interactions were taken into account \cite{ee1,ee2}. However, such analysis
were centered in the undoped (half-filling) regime or very close
to it, and explicit analytic results for a wide range of
fillings are still lacking.

In the present paper we apply the method developed in \cite{LCG} to
fermions in a graphene sheet in the presence of electron-electron
interactions up to nearest neighbors. A previous work using a mean
field approach, showed that long range interactions could lead to
Pomeranchuk instabilities \cite{Belen} at the Van Hove filling. On
the other hand, at dopings very close to half-filling, it was anticipated that
graphene should behave as a Fermi liquid \cite{Sarma}. This was later confirmed
experimentally by ARPES exploration of the FS \cite{experimentales}.
The results presented here are consistent with these findings, while our
method allows for a more complete and systematic study of the whole
space of fillings.

\section{Two dimensional Pomeranchuk instabilities: review and improvement of the method}

In this section we will review the generalization of Pomeranchuk method first presented in \cite{LCG}, and discuss a shortcut that can be used as an alternative to the change of variables proposed there.

\subsection{Review of the method}
\label{sec:proof}
According to Landau's theory of the Fermi liquid, the free-energy $E$ as a functional of the change $\delta n_\vk$ in the equilibrium distribution function at finite chemical potential $\mu$ can be written, to first order in the interaction, as

\small
\ba
&&\!\!\!\!\!\!\!\!\!\!\!\!
 E\!=\!\int \!\!d^{2}\!\vk\,(\varepsilon(\vk)\!-\!\mu)\,\delta n_\vk 
 +\frac{1}{2}\!\int \!\!d^{2}\!\vk \!\!\int\! \!d^{2}\!\vk'
f(\vk,\vk')\;\delta n_\vk\delta n_{\vk'} \, .\ \     
\label{deltaE}
\ea
\normalsize
Here $\varepsilon(\vk)$ is the dispersion relation that controls the free dynamics of the system,
the interaction function $f(\vk,\vk')$ can be related to the low
energy limit of the two particle vertex. Note that we are omitting spin indices, and considering only variations of the total number of particles $n_\vk=n_{\uparrow \vk}+n_{\downarrow\vk}$. This implies that the considerations that follow will be valid in the absence of any external magnetic field and at constant total magnetization. The identification of these two functions is the starting step of our calculation:

\vspace{0.5cm}

\step{1}{write the energy as in (\ref{deltaE}) and identify the functions $\varepsilon(\textbf{k})$ and $f(\textbf{k},\textbf{k}')$.}

Instead of the cartesian variables in momentum space $(k_x,k_y)$ we will find convenient to define a new set of curvilinear  variables $(g,s)$ according to
\ba
g&=&g(\textbf{k})\equiv\mu-\varepsilon(\textbf{k})\,,\nonumber \\
s&=&s(\textbf{k})\, . \label{changeofvariables}
\ea
The variable $g$ varies in the direction normal to the Fermi Surface (FS), whose position is defined by $g=0$.
We choose $s$ such that it is constant at constant distance to the FS, varying in the longitudinal direction tangent to the FS, namely it satisfies the restriction
\be\nabla s(\textbf{k}).\nabla g(\textbf{k})=0\,.\label{constraintcomplicado}\ee
Since solving this for $s$ may be a difficult task, we develop bellow an alternative procedure that replaces this calculation. Even if not mandatory, we will chose the variable $s$ such that we give a complete turn around each connected piece of the FS when it runs from $-\pi$ to $\pi$.

From the above change of variables we obtain the Jacobian
\be
J^{-1}(g,s)=\left|\frac{\partial(g,\,s)}{\partial \textbf{k}}\right|\,,
\label{jaco}
\ee
which is the relevant outcome of this step of the calculation:

\vspace{0.7cm}

\step{2}{with the help of the dispersion relation $\varepsilon(\textbf{k})$ obtained in step 1, change the variables according to (\ref{changeofvariables}) to obtain the Jacobian (\ref{jaco}).}

In a stable system, the energy (\ref{deltaE}) should be positive for all $\delta n_\vk$ that satisfy the constraint imposed by Luttinger theorem \cite{Luttinger}
\be
\int d^2\! \vk \,\delta n_\vk = 0\,.
\label{eq:luttinger}
\ee

Pomeranchuk's method roughly consists on exploring the space of solutions of constraint (\ref{eq:luttinger}) to find a $\delta n_\vk$ that turns the energy into negative values, thus pointing to an instability of the system.

In terms of our new variables $g$ and $s$, we can write $\delta n_{\vk(g,s)}$ at zero temperature as
\ba
\delta n_{\vk(g,s)}&=&H[g +\delta g(g,s)]-H[g]\, ,
\label{eq:deltaN}
\ea
where $H$ is the unit step function and $\delta g(g,s)$ is an small perturbation parameterizing the deformation of the FS. Replacing into the constraint (\ref{eq:luttinger}), changing the variables of integration according to (\ref{changeofvariables}) and performing the integral to lowest order in $\delta g$ we get
\be
\int ds J(s)\delta g(s)=0\,.
\label{contraint}
\ee
Here $J(s)=J(g,s)|_{g=0}$ and $\delta g(s) = \delta g(g,s)|_{g=0}$. In case the FS has a nontrivial topology, the integral includes a sum over all different connected pieces.

We see that in order to solve the constraint
$\delta g(s)$ can be written as
\ba
\delta g(s)
\simeq
J^{-1}(s)\partial_s \lambda(s)\,,
\label{eq:lambda}
\ea
in terms of a free slowly varying function $\lambda(s)$. Even if in (\ref{eq:luttinger}) and (\ref{contraint}) a sum over different connected pieces of the FS may be assumed, this particular solution does not consider excitations in which some particles jump  between different connected pieces .

~

Using the change of variables (\ref{changeofvariables}) and with the help of eqs. (\ref{eq:deltaN}) and (\ref{eq:lambda}), we can rewrite the energy $E$ to lowest order in $\delta g$ as
\ba E\!\!&=&\!\!\int \!ds'\!\int \!ds \,\psi(s')\frac12
\left(\phantom{\frac12}\!\!\!\! J^{-1}(s)\delta(s-s') + f(s,s')
\right)\psi(s)\,,
\nonumber\\&&
~\!\!\!\!\!\!
\!\!\!\!\!\!\!\!\!\!\!\!\!\!\!\!\!\!
\label{bilinear}
\ea
where we call  $\partial_s\lambda(s)=\psi(s)$ and
$ f(s,s')=f(g,s;g',s')|_{g=g'=0}$.

Note that the left hand side of the stability condition $E>0$ has two terms, the first of which
contains the information about the form of the FS via $J^{-1}(s)$,
while the second encodes the specific form of the interaction in
$f(s,s')$. There is a clear competition between the interaction
function in the second term of the integrand and the first term that
only depends on the geometry of the unperturbed FS.

We see in (\ref{bilinear}) that the energy $E$ is a bilinear form acting on the real functions $\psi(s)$ that parameterize the deformations of the FS
\ba
E=\langle\psi,\psi\rangle\,.
\ea
This is the next step of our calculation that can be summarized as:

\vspace{0.4cm}

\step{3}{write the energy as the bilinear form (\ref{bilinear}) using the functions $J(s)$ and $f(s,s´)$ identified in steps 1 and 2.}

The stability condition is then equivalent to demanding this bilinear form to be
positive definite for any possible smooth deformation of the FS
\ba
\forall\psi\mbox{\small$\in L_2[\mbox{FS}]$\normalsize}: \,\langle\psi,\psi\rangle>0\,,
\ea
where $L_2[\mbox{FS}]$ stands for the space of square-integrable functions defined on the Fermi surface.

In consequence, a straightforward way to diagnose an instability is to
diagonalize this bilinear form looking for negative eigenvalues.
To see that, we choose an arbitrary basis of functions $\{\psi_n\}_{n\in N}$ of $L_2[\mbox{FS}]$, in terms of which we can write
\be
\psi(s) = \sum_n a_n\psi_n(s)\,,
\ee
and the the stability condition now reads
\be
E=\sum_{n,m}a_na_m^*\langle\psi_n,\psi_m\rangle >0\,.
\ee
 This configures our

 \vspace{0.7cm}

\step{4}{choose an arbitrary basis $\{\psi_n\}_{n\in N}$ of the space of functions $L_2[\mbox{FS}]$.}

The above defined bilinear form can be considered as a pseudo-scalar product in $L_2[\mbox{FS}]$. In general the functions of the basis $\{\psi_n\}_{n\in N}$ will not be mutually orthogonal with respect to this product. Moreover in the presence of instabilities, the pseudo-scalar product may lead to negative pseudo-norms $\langle\psi,\psi\rangle <0$.

We can then make use of the Gram-Schmidt orthogonalization procedure to obtain a new basis of mutually orthogonal functions $\{\xi_n\}_{n\in N}$ . In terms of them an arbitrary deformation of the FS parameterized by a function $\psi(s)$ can be decomposed as
\ba
\psi(s)=\sum_{n}\;b_{n}\xi_{n}(s)\,,
\ea
which implies that the stability condition on such deformation will read
\ba
\label{eq:funcional diagonal}
E=\sum_{n}\;|b_{n}|^{2}\;\langle \xi_{n},\xi_{n}\rangle>0\,.
\ea
In summary:

\vspace{0.7cm}

\step{5}{apply the Gram-Schmidt orthogonalization procedure to go from the arbitrary basis $\{\psi_n\}_{n\in N}$ chosen on step 3 onto a basis of mutually orthogonal functions $\{\xi_n\}_{n\in N}$.}

In (\ref{eq:funcional diagonal}) we note that the only possible source of a negative sign is in the pseudo-norms $\chi_{n}=\langle \xi_{n},\xi_{n}\rangle$. In case the $i$-th pseudo-norm $\chi_i$ is negative, a deformation parameterized by $\psi(s)\propto\xi_i(s)$ is unstable. In other words the pseudo-norms $\{\chi_{n}\}_{n\in N}$ can be taken as the
stability parameters, a negative value of $\chi_{i}$ indicating
a instability in the $i$-th channel. Then:

\vspace{0.7cm}

\step{6}{compute the pseudo-norms of the new basis functions $\{\xi_n\}_{n\in N}$. If for a given channel $i$ one finds that $\chi_{i}=\langle\xi_i,\xi_i\rangle<0$, the
FS is diagnosed to be unstable.}

These six steps summarize the generalized Pomeranchuk method. It can be applied to any two dimensional model with arbitrary dispersion relation and interaction. Note that since $L_2[\mbox{FS}]$ is infinite dimensional, the present
method is not efficient to verify stability: at any step $i$ it may
always be the case that for some $j$, $\chi_{i+j}<0$. Moreover, in the case of nontrivial topology, excitations consisting on particles jumping between different connected pieces of the FS were not included in the solution of the constraint (\ref{contraint}), and they may lead to additional instabilities. The same is true excitations involving spin or color flips, that we are not considering.


\subsection{An alternative to the change of variables}

As advanced, to avoid the task of solving the constraint (\ref{constraintcomplicado}) that defines the variable $s$, we will develop here an alternative procedure to derive the form of the Jacobian evaluated on the FS $J(s)$.

We begin by defining a parametrization of the FS
\ba
\label{eq:parametric}
\vsigma(t)=
\left(k_x(t) , k_y(t) \right)\,, \hspace{1cm}   -\pi <t<\pi\,,
\ea
in terms of an arbitrary parameter $t$. In other words, given the function $g(\textbf{k})$ defined in (\ref{changeofvariables}), we choose $\vsigma(t)$ such that $\forall t: g(\vsigma(t))=0$. In terms of such parametrization we can decompose the Dirac $\delta$ function as
\be
\delta(g(\textbf{k}))= \int dt\; \frac{|\dot{\vsigma}(t)|}{|\mathbf{\nabla }g(\vsigma(t))|}\,\delta\!^{(2)}\!(\textbf{k} - \vsigma(t))\,,\label{delta}
\ee
(a proof of this formula is given in the Appendix).

The integral $I$ of an arbitrary function $F(g,s)$ along the FS can be written as
\be
I=\int ds \left.F(g,s) \right|_{g=0}\,,
\label{integral}
\ee
or in other words
\be
I=\int ds\,dg \,F(g,s) \,\delta(g)\,.
\ee
Changing variables to $\textbf{k}$
\small
\ba
I&=&\int d^2\vk\,F(s(\textbf{k}),g(\textbf{k}))
\;J^{-1}\!\!\left(s(\textbf{k}),g(\textbf{k})\right)\, \delta(g(\textbf{k}))\,,\ \ \ \
\ea
\normalsize
and replacing (\ref{delta}) we get
\small
\ba
I&=&\int d^2\vk\,F(s(\textbf{k}),g(\textbf{k}))\;J^{-1}\!\!\left(s(\textbf{k}),g(\textbf{k})\right)\,
\times\nonumber\\&&\ \ \ \ \ \ \times
\int dt \frac{|\dot{\vsigma}(t)|}{|\mathbf{\nabla }g(\vsigma(t))|}\,\delta\!^{(2)}\!(\textbf{k} - \vsigma(t))\,,
\ea
\normalsize
or, interchanging the order of the integration and performing the $\vk$ integrals
\small
\ba
I\!&=&\!\int \!\!dt \,F(s(\vsigma(t)),g(\vsigma(t)))\;J^{-1}\!\!\,\left(s(\vsigma(t)),g(\vsigma(t))\right)
\,
\frac{|\dot{\vsigma}(t)|}{|\mathbf{\nabla }g(\vsigma(t))|}\,.\nonumber\\&&
\ea
\normalsize
Now using the fact the $g(\vsigma(t))=0$ and defining the parameter $t$ such that $s(\vsigma(t))=t$, we get
\ba
I&=&\int \!ds \,\left.F(s,g)\right|_{g=0}\,J^{-1}(s)\,
\frac{|\dot{\vsigma}(s)|}{|\mathbf{\nabla }g(\vsigma(s))|}\,,
\ea
that in order to be compatible with (\ref{integral}) imply
\be
J^{-1}(s)=\frac{|\mathbf{\nabla }g(\vsigma(s))|}{|\dot{\vsigma}(s)|}\,.
\label{jac}\ee

Then with all the above, we can replace our previous step 2 by a new version
\vspace{0.7cm}

\step{2'}{with the help of the dispersion relation $\varepsilon(\textbf{k})$ obtained in step 1, parameterize  the FS and obtain the Jacobian evaluated on the FS according to (\ref{jac}).}

Then, even if it may be very difficult to solve the partial differential equation (\ref{constraintcomplicado}) in order to explicitly obtain the Jacobian, its restriction to the FS is all what we need, and can be obtained by the much easier task of parameterizing the FS.

\section{Pomeranchuk instability in graphene}

In the present section we will apply the method reviewed in the previous pages to the case of fermions in a graphene layer with Coulomb interactions.

\subsection{Free Hamiltonian: tight-binding approach}

Graphene is made out of carbon atoms arranged in a hexagonal lattice. It is not a Bravais lattice but can be seen as a triangular lattice with two atoms per unit cell. The tight-binding Hamiltonian for electrons in graphene considering
that electrons can hop only to nearest neighbor atoms has the form

\be H_0=-t \sum_{\langle i,j \rangle,\sigma}
\left(a_{\sigma i}^{\dag} b_{\sigma j} + {\rm H.c.} \right) \,,
\ee
where $a_{\sigma i}, b_{\sigma i}$ are the creation and annihilation operators related to each of the unit cell atoms.

The diagonalized Hamiltonian can be written in terms the
occupation numbers of rotated lattice operators defined by
\be n^{\!\pm}_{\sigma \vk} = \frac{1}{2}\!\left(b_{\sigma\vk}^\dagger \pm
\frac{h^*(\vk)}{\varepsilon(\vk)} \,a_{\sigma\vk}^\dagger \right)\left(b_{\sigma\vk} \pm
\frac{h(\vk)}{\varepsilon(\vk)} \,a_{\sigma\vk} \right)\,,
\ee
where the function $h(\vk)$ satisfies $\vert h(\vk) \vert^2=\varepsilon(\vk)^2$ and reads
\ba
h(\vk)\!\!&=&\!\!t
\left(\cos(k_{x})  -2i\sin(k_{x}\!-\!1) + 2\cos^{2}(\frac{k_{x}}2)  +\phantom{\frac{\sqrt2}2}\!\!\!\!\!
\right.
\nonumber\\&&
\left. +4i\cos(\frac{\sqrt{3}k_y}2)\sin(\frac{k_{x}}2) +4\cos(\frac{\sqrt{3}k_y }2) \cos(k_{x})
\right)\,.
\nonumber\\
\ea
It results in
\be
H_0 = \int d^2\vk\sum_\sigma
\left(
\varepsilon_{\sigma}^+(\vk) n_{\sigma\vk}^+ + \varepsilon_{\sigma}^-(\vk) n_{\sigma\vk}^-
\right)\,,
\label{Hfree}
\ee
here we defined the energy bands by $\varepsilon_{\sigma}^\pm(\textbf{k})=\pm\varepsilon(\textbf{k})$ with
\small
\ba
\varepsilon(\vk)\!\!&=&\!\!t\sqrt{3\!+\!4 \cos \left(\frac{3 }{2}k_x a\right) \cos \left(\frac{\sqrt{3} }{2}k_y a\right)\!+\!2 \cos \left(\sqrt{3} k_y a\right)}\,,
\nonumber\\&&
\label{dispersion}
\ea
\normalsize
where $a$ is the carbon-carbon distance ($a \sim 1.24 $ {\AA}).

The energy resulting from a small variation on the occupation numbers on (\ref{Hfree}) at finite chemical potential $\mu$  reads
\be
E_0 = \int d^2 \!\vk\,(\varepsilon(\vk)-\mu) (\delta n_{\vk}^+ - \delta n_{\vk}^-)\,,
\label{Efree}
\ee
where use have been made of the fact that the dispersion relation do not have spin indices to write the resulting expression in terms of variations of $n_\vk^\pm=n^\pm_{\uparrow\vk}+n^\pm_{\downarrow\vk}$.

To fix the ideas we consider only non-vanishing variations of the
occupation numbers in the upper band, {\it i.e.} $\delta n^-_k=0$,
$\delta n^+_k \ne 0$. This can be done without loss of generality due to the symmetry of the system under the interchange of the positive and negative bands. We then have
\be
E_0 = \int \!d^2 \vk \,(\varepsilon(\vk)-\mu) \delta n_{\vk}^+\,.
\ee
This has the form o the free term of (\ref{deltaE}), giving us one of the ingredients required by our step 1 defined above, namely the dispersion relation $\varepsilon(\vk)$.

\begin{figure}[t]
  \includegraphics[width=260pt]{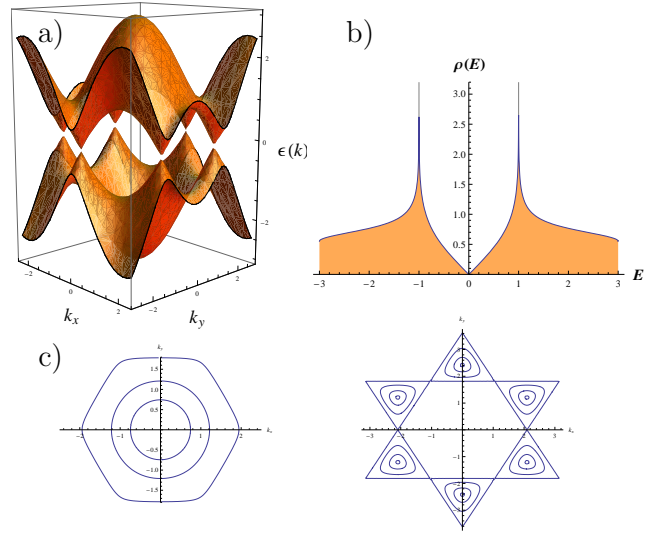}\\
  \caption{(Color online) a) Energy spectrum for the tight-binding approach. b) Density of states per unit cell as a function of the energy. All the quantities are given in units of $t$. c) Left: FS for $t<\mu<3t$. Right: FS for $0<\mu<t$ }\label{fig:intro}
\end{figure}

The band structure  is shown in Fig \ref{fig:intro}a. The density of states is presented in Fig. \ref{fig:intro}b where a Van Hove singularity can be seen to be present for $\mu=\pm t$. The resulting FS's for different fillings are shown in Fig \ref{fig:intro}c.

%


\subsection{Interactions in graphene}
\label{sec:Hubbard}


To complete the ingredients required by our step 1, we need the
quasiparticle interaction function $f(\vk,\vk' )$. In what follows,
for completeness and to set up our conventions, we briefly describe
how to derive its expression to first order in a perturbative
expansion \cite{Abrikosov,Quintanilla1}.

We will consider density-density interactions, both on-site (with
strength $U$) and between nearest-neighbors (with strength $V$),
namely our interaction Hamiltonian reads
\ba
H_{int}=\frac U2\sum_{i}n_{i}(n_{i}-1) +
V \sum_{\langle i,j\rangle}n_{i }n_{j}\,,
\ea
where $\langle i,j\rangle$ stands for nearest neighbors, and the density operators $n_{i }= n_{i \uparrow} + n_{i \downarrow}$ refer to the original lattice operators $a_i,b_i$.

One then computes the energy in the mean field approximation. The
result can be written in terms of the mean field values of the
occupation numbers of the rotated lattice operators that diagonalize $H_0$, reading
%
%
%
%
%
\small \ba \langle H\rangle_{\!M\!F}\!\!
&=&
\!\!\frac{1}{2N}\!\!\!\!\!\sum_{\vk,\vk',\alpha,\beta}
\!\!\!\!\!\left(n^{\!-}_{\vk\alpha}n^-_{\vk'\beta} \!+\! n^{\!+}_{\vk\alpha}
n^{\!+}_{\vk'\beta} \!+\!  n^{\!+}_{\vk\alpha} n^{\!-}_{\vk'\beta}\!+\!
n^{\!-}_{\vk\alpha} n^{\!+}_{\vk'\beta}\right) \!\!\times
\nonumber \\&&
\ \ \times
\left(U\sigma^1_{\alpha\beta} + \frac{V}{2}\left( {F(0)}
(\sigma^0_{\alpha\beta}\!+\!\sigma^1_{\alpha\beta})\right)\right)-
\nonumber \\&-&\!\!
\frac{1}{2N}\!\!\!\!\sum_{\vk,\vk',\alpha,\beta}
\!\!\!\!\!\left(n^{\!-}_{\vk\alpha}n^-_{\vk'\beta} \!+\! n^{\!+}_{\vk\alpha}
n^{\!+}_{\vk'\beta} \!-\!  n^{\!+}_{\vk\alpha} n^{\!-}_{\vk'\beta}\!-\!
n^{\!-}_{\vk\alpha} n^{\!+}_{\vk'\beta}\right)
\!\!\times \nonumber \\&&\ \ \times
\frac{V}{2}\left(
\sigma^0_{\alpha\beta}F(\vk-\vk')\frac{\varepsilon(\vk)\varepsilon(\vk')}{h^*(\vk)h(\vk')}\right)\,,
\ea \normalsize
where $N$ is the number of sites and
\be
F(\vk)=\sum_{\alpha =1}^3e^{i\vk\cdot \mathbf{\delta}_{\alpha}}\,,
\ \ \ \ \ \ \ \mbox{with} \ \ \ \ \ \
\left\{
\begin{array}{l}
\mathbf{\delta_1}=a(\frac{1}{2},\frac{\sqrt{3}}{2})\,,\\
\mathbf{\delta_2}=a(\frac{1}{2},-\frac{\sqrt{3}}{2})\,,\\
\mathbf{\delta_3}=a(-1,0)\,.
\end{array}
\right.
\ee

As advanced in the previous section, we will concentrate in variations of the occupation numbers that keep the total magnetization constant. In other words, we assume $\delta n^+_{\uparrow\vk} = \delta n^+_{\downarrow\vk}$. Similarly to the free part, the
interactions between quasiparticles can then be written in terms of the variation of the total number $n^+_\vk=n^+_{\uparrow\vk}+n^+_{\downarrow\vk}$ as
\be
E_{int}=\int \!d\vk d\vk'\, f(\vk,\vk')\delta n^+_\vk \delta n^+_{\vk'}\,.
\ee
The function $f(\vk,\vk')$ is then obtained from the mean
field value of the energy 
\ba f(\vk,\vk')\!\!&\equiv& \!\!\frac{\delta^{(2)}\! \langle
H\rangle_{\!M\!F}}{\delta n^+_\vk\delta
n^+_{\vk'}} =\nonumber \\
&=&\!\! \frac 1{2(2\pi)^2} \!\! \left(\!U \!+\! \frac
V2\!\left({F(0)}\!-\!
{F(\vk\!\!-\!\!\vk')}\frac{\varepsilon(\vk)\varepsilon(\vk')}{h^{*}(\vk)h(\vk')}\right)\!\right)\,.
\nonumber\\
\label{interaction}\ea

~

We have then completed step 1, getting the dispersion relation (\ref{dispersion}) and the interaction function (\ref{interaction}).

\subsection{Parametrization of the Fermi surface}

Step 2' requires the parametrization of the FS, for which we need to study
separately fillings lying above and below the Van Hove filling.
In what follows we present the parameterized curves used throughout the calculations.

\subsubsection{High energy sector: $|\mu | > t$}

We call high-energy sector the case in which $t<|\mu|<3t$, {\em i.e.} the region of fillings which lie above the Van Hove singularity. As can be seen in Fig.\ref{fig:intro}c, for $\mu /t \sim 3$ the FS is approximately circular, while for values closer to 1 the FS takes a hexagonal form.

In this sector the FS can be parameterized as follows
\ba
\label{eq:parametricH}
\vsigma^{H}(s)=\left(\;  k_x^{H}(s) \; , \;  k_y^{H}(s) \; \right) \,,\hspace{1cm}   -\pi <s<\pi\,,
\ea
where
\ba
\label{eq:parametric2}\nonumber
k_x^{H}(s)&=&\frac{2}{3\,a}\text{sign}(s)\,\arccos \!\left[G(k_y^{H}(s)) \right]\,,\\
k_y^{H}(s)&=&\frac{1}{\sqrt{3}a}(\phi_{H}-\omega_{H}\;|s|)\,,
\ea
with $\omega_{H}$, $\phi_{H}$ and the auxiliary function $G(x)$ defined as
\ba\label{eq:parametric3}\nonumber
&&\!\!\!\!\!\!\!\!\!\!\!\!\!\!\!\omega_{H}=\frac{4}{\pi } \arccos\!\!\left(\frac{\mu \!-\!t}{2 t}\right)\,,\ \ \
\phi_{H}=2 \arccos\!\!\left(\frac{\mu\!-\!t }{2 t}\right)\,,\\
&&\!\!\!\!\!\!\!\!\!\!\!\!\!\!\!G(x)=\frac{1}{4} \left(\frac{\mu ^2}{t^2}-2 \cos(\sqrt{3} \,x)-3\right) \sec \!\left(\!\frac{\sqrt{3} \,x }{2}\right)\,.
\ea

\subsubsection{Low energy sector: $|\mu | < t$}

The low energy sector corresponds to fillings satisfying $0< |\mu| <t$. In this case the FS consists of pockets centered at the vertices of the Brillouin zone, as can be seen in Fig.\ref{fig:intro}c. By using the periodic identifications of the momentum plane, we see that only two of them are non-equivalent. In consequence one can describe the total FS as two pockets centered in the two Dirac points $\textbf{k}_{\pm}=(0,\pm \frac{4\pi}{3a})$.

For example, for the FS pocket centered in $(0,\frac{4\pi}{3a})$  we can
choose a parametrization of the form
\ba
\label{eq:parametricL}
\vsigma^{L}(s)=\left(\;  k_x^{L}(s) \; , \;  k_y^{L}(s) \; \right) \,,\hspace{1cm}   -\pi <s<\pi\,,
\ea
with
\ba\label{eq:parametric4}\nonumber
k_x^{L}(s)&=&\frac{2}{3\,a}\text{sign}(s)\,\arccos \!\left[G(k_y^{L}(s)) \right]\,,\\
k_y^{L}(s)&=&\frac{1}{\sqrt{3}a}(\phi_{L}-\omega_{L}\;|s|)\,,
\ea
and
\ba
\label{eq:parametric5}\nonumber
&&\!\!\!\!\!\!\!\!\!\!\!\!\!\!\!\!\omega_{L}=\frac{2}{\pi } \left(\arccos\!\!\left(\frac{-t\!-\!\mu }{2 t}\right)-\arccos\!\!\left(\frac{\mu \!-\!t}{2 t}\right)\right)\,,
\\
&&\!\!\!\!\!\!\!\!\!\!\!\!\!\!\!\!\phi_{L}=2 \arccos\!\!\left(\frac{-t\!-\!\mu }{2 t}\right)\,.
\ea

~

With the above parametrizations of the high and low energy sectors, we can compute
the corresponding Jacobian evaluated on the FS according to (\ref{jac}), obtaining

\begin{widetext}
\ba
J^{-1}(s)
&=&\left\{
\begin{array}{l}
\frac{3\sqrt{3}}{4|\mu | |\omega_{H}|}
   \sqrt{6 \mu ^2-\mu ^4+4 \left(\mu ^2-1\right) \cos \left( \omega_{H}  (\frac{\pi}{2} \!-\! |s|)\right)-2
   \cos (\omega_{H}  (\pi\! -\!2 |s|))-3}\,,
   \ \ \ \ \ \ \ t<|\mu|<3t\,,
\\
\\
\frac{3\sqrt{3}}{4 {|\mu | |\omega_{L}|}}
   \sqrt{\phantom{\frac\pi2}\!\!\!\!\!\,6 \mu ^2-\mu ^4+4 \left(\mu ^2-1\right) \cos \left(  \phi_{L}  \!- \!  \omega_{L}  |s|\right)
   -2 \cos \left(2\left(  \phi_{L}\!  -\!   \omega_{L}  |s|\right)\right)-3}\,,
  \ \ \ \ \ \ \ 0<|\mu|<t\,.
\end{array}
\right.
\ea
\end{widetext}
This completes our step 2', providing us with the values of the Jacobian evaluated on the FS $J^{-1}(s)$. The interaction function evaluated on the FS $f(s,s')$ is obtained by simply replacing the parameterizations of the high and low energy sectors in (\ref{interaction}). This allows us to complete step 3, by constructing the energy function as a bilinear form.

\subsection{Instabilities and phase diagram}

To proceed with step 4, we choose a basis of the space of functions $L_{2}[\mbox{FS}]$. Here for simplicity we choose trigonometric functions
\be
\{\psi_n(s)\}_{n\in N}=\{\cos(ms),\sin(ms)\}_{m\in N}\,.
\ee
\begin{figure}[t]
  \includegraphics[width=220pt]{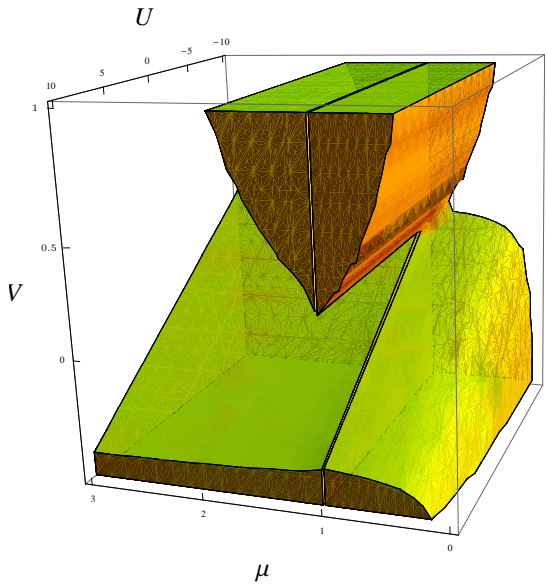}\\
  \caption{(Color online) 3D Phase diagram for graphene. The phase diagram was constructed by exploring the first 20 modes, the shaded region is unstable. The first three modes cover most of the instability region, the remaining modes just re-draw the details of the boundary. Note that for purely repulsive interaction (positive $U$ and $V$) there is an unstable region near the Van Hove filling $\mu\simeq 1$. On the other hand, near half fillings $\mu\simeq 0$ the Fermi liquid is stable for any value of the interactions.\label{fig:1}}
\end{figure}

Following step 5 by means of an orthogonalization procedure, we obtain the new basis of mutually orthogonal functions $\{\xi_n\}_{n\in N}$.

In terms of this new basis and according to step 6, we compute the stability parameters $\chi_{i}=\langle\xi_i,\xi_i\rangle$.
These are functions on the space of parameters $(\mu, U, V)$ that may become negative in some regions. If this is the case, in such regions the Fermi liquid is diagnosed to be unstable.

In our calculations, this last step was performed numerically, due to the complication of the integrals involved in the pseudo-norms $\{\chi_n\}_{n\in N}$.

~

\section{Results}

The phase diagram on the space spanned by the interaction strengths $U$, $V$ and the chemical potential $\mu$ is shown in Fig \ref{fig:1}. There we plot the instability region determined by the dominant unstable channels.

The method used in this work allows to explore all possible fillings and to draw a phase diagram for graphene systems valid both below and above half-filling. The advantage of our approach lies in the fact that it can be pursued systematically following the steps described in previous sections, studying separately each deformation mode of the FS.

For fillings around the Van Hove filling the Pomeranchuk instability is favored. For on-site and nearest-neighbor Coulomb repulsion ($U>0$ and $V>0$) we find a region of parameter space where the system presents an instability. Near to the Van Hove filling our results are in agreement with those found using a mean field approach in \cite{Belen}.

On the other hand,  we do not see any instability around half filling. This is in agreement with the results presented by Sarma {\it et al} in \cite{Sarma} for doped graphene, where the authors found that extrinsic graphene is a well defined Fermi liquid for low energies, within the Dirac fermion approximation. Moreover, this agrees with experimental results found using ARPES presented in \cite{experimentales}, implying that graphene is a Fermi liquid for low dopings.

\begin{figure}[t]
  \includegraphics[width=240pt]{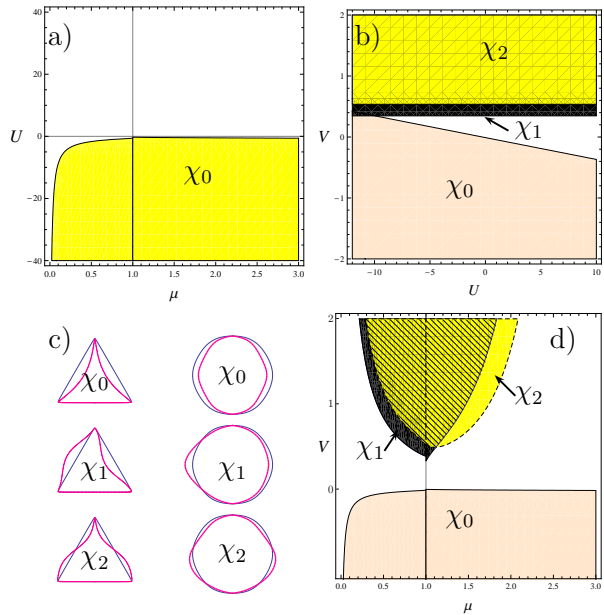}\\
  \caption{(Color online) Instability sectors for the first unstable modes. $a)$ Instability for the $\chi_{0}$ mode in the $V=0$ plane. It is reached only for attractive on-site interactions ($U<0$). $b)$ Instability  regions on the plane $\mu=0.99$: the modes $\chi_{1}$ and $\chi_{2}$ are unstable for  $U>0$ and $V>0$. $c)$ FS deformations corresponding to the first unstable modes: the left column corresponds to the low energy sector and right column to the high energy one. $d)$ Instability reached by the $\chi_{0}$, $\chi_{1}$ and $\chi_{2}$ modes in the $U=0$ plane. Remarkably, it is present for repulsive interactions ($U>0$,$V>0$).
  }\label{fig:planos}
\end{figure}

For attractive on-site interaction, the region where the instability is detected depends strongly of the nearest neighbors interaction strength.

We find that even the smoothest deformations of the FS, {\em i.e.} those described by lower modes in our orthogonal basis, present instabilities. Indeed, they cover most of the unstable region. In Fig.\ref{fig:planos}  instability regions corresponding to the first modes are drawn. In Fig.\ref{fig:planos}a the $V=0$ plane is shown, the unstable mode corresponding to the colored region is the 0-th mode and the corresponding FS deformation is presented in Fig.\ref{fig:planos}c. In Fig.\ref{fig:planos}b and \ref{fig:planos}d the instability regions for the first modes are plotted in the $U$-$V$ and $\mu$-$V$ plane respectively. In both Figures the $\chi_{0}$ instability can be seen, and a new region appears where an instability of the $\chi_{1}$ and $\chi_{2}$ channels is present. This instability appears for positive values of the interaction strength and is closer to the Van Hove filling. This is in agreement with earlier results, where a Pomeranchuk instability at the Van Hove filling was found within a mean field approximation \cite{Belen}. The interacting FS presented there has the same geometry than that of the corresponding deformation channel $\chi_{2}$ shown in the left column of Fig.\ref{fig:planos}c.

The results presented above confirm earlier predictions about the Fermi liquid behavior of graphene for the cases near Van Hove filling and near half filling. They also provide a more complete description of the phase diagram for the entire range of fillings in between these two limiting cases.

\section{Summary and Outlook}
\label{sec:conclusions}

We have explored Pomeranchuk instabilities in graphene using a recently developed generalization of Pomeranchuk method. We have obtained the phase diagram of the dominant instability as a function of the on-site $U$ and nearest neighbor $V$ interaction strengths and the chemical potential $\mu$ (Fig.\ref{fig:1}). We analyzed several planes of the 3D phase diagram obtaining a good agreement with previous theoretical results and experimental findings (Fig.\ref{fig:planos}).

The phase diagram makes apparent some interesting features of the system. For example no instability is detected at low energies. This behavior is noteworthy because in this sector a Dirac massless fermions approach can be used to describe the graphene layer in the absence of interactions. On the other hand, for energies close to the  Van Hove energy, instabilities appear to cover a large region in the $U$-$V$ plane.

\vspace{1cm}

The efficiency of the method shows up in the fact that the complete phase diagram is obtained by exploring a few number of modes. The introduction of higher modes does not enlarge substantially the region of instability but re-draw the details of its boundaries.

The method can be applied either analytically or numerically according to the complexity of the system under investigation.
In the previously studied case \cite{LCG}, the analytical calculations were pursued up to the end, allowing us to draw the phase diagram. In the present case, the calculations were performed analytically up to the point of the evaluation of the instability parameters $\chi_i$, which involved complicated integrals that were then performed numerically. The method is suitable for direct application to numerical data encoding the form of the Fermi surfaces, like those obtained by application of ARPES.

The form of the method presented here is suitable for any two-dimensional lattice model at zero temperature. However, it does not consider instabilities arising from particles jumping between different disconnected pieces of the FS or from spin or color flipping. It can be easily extended to consider these effects, as well as to three dimensional systems, such as multilayer graphene, ruthenates, etc. The generalization to finite temperatures involves a different definition of the pseudo scalar product. All these extensions will be presented in a forthcoming work \cite{wip}.

\section{Appendix}

Given the implicit curve defined by $g(\textbf{k})=0$, we can choose a parametrization in terms of a vector function
\ba
\label{eq:parametricApp}
\vsigma(t)=
\left(k_x(t) , k_y(t) \right) \,,\hspace{1cm}   -\pi <t<\pi\,,
\ea
depending on an arbitrary parameter $t$, and defined so that $\forall t:\,g(\vsigma(t))=0$.
In terms of this parametrization we want to prove that the Dirac $\delta$ function can be written as
\be
\delta(g(\textbf{k}))= \int dt \frac{|\dot{\vsigma}(t)|}{|\nabla g(\vsigma(t))|}\,\delta\!^{(2)}\!(\textbf{k} - \vsigma(t))\,.
\label{delta-1}
\ee
To that end, we write more explicitly the right hand side
\be
\delta(g(\textbf{k}))= \int dt \frac{|\dot{\vsigma}(t)|}{|\nabla g(\vsigma(t))|}\,
\delta(k_x - k_x (t))\,\delta(k_y - k_y (t))\,,
\label{delta_app}
\ee
and then rewrite the $k_x$ delta function using the well known one dimensional formula
\be
\delta(f(x))=\frac{\delta(x-\bar x)}{f'(\bar x)}\ \ \ \ \ \mbox{where}\ \ \ \ f(\bar x)=0\,,
\label{delta-2}
\ee
to get
\be
\delta(g(\textbf{k}))= \int dt \frac{|\dot{\vsigma}(t)|}{|\nabla g(\vsigma(t))|}\,
\frac{\delta(t-t(k_x))}{\dot k_x(t)}
\,\delta(k_y - k_y (t))\,,
\label{delta-3}
\ee
where we call $t(k_x)$ to the solution of $k_x-k_x(t)=0$. Performing the $t$ integral
\be
\delta(g(\textbf{k}))= \frac{|\dot{\vsigma}(k_x)|}{|\nabla g(k_x)|}\,
\frac {\delta(k_y - k_y(k_x))}{\dot k_x(k_x)}\,,
\label{delta-4ymedio}
\ee
where we use the notation $f(k_x)=f(t(k_x))$ for any function $f$. Writing explicitly the square roots on the vector norms
\small
\ba\
\delta(g(\textbf{k}))=\frac{\sqrt{{\dot k_x}^2\!(k_x)+{\dot k_y}^2\!(k_x)}\,\delta(k_y - k_y(k_x))}{\sqrt{(\partial_xg(k_x,k_y(k_x)))^2\!+\!(\partial_yg(k_x,k_y(k_x)))^2}\,\dot k_x(k_x)}\,.
\nonumber\\
\ea
\normalsize
A further rearrangement of the formulas gives
\be
\delta(g(\textbf{k}))= \frac{\sqrt{1+\left(\frac {dk_y(k_x)}{dk_x}\right)^2}\,\delta(k_y - k_y (k_x))}{\sqrt{1+\left(\frac{\partial_xg(k_x,k_y(k_x))}{\partial_yg(k_x,k_y(k_x))}\right)^2}\,\partial_yg(k_x,k_y(k_x))}\,,
\ee
where we can identify the derivative in the numerator with the quotient of derivatives in the denominator to cancel the square roots, obtaining
\be
\delta(g(\textbf{k}))= \frac{\delta(k_y - k_y (k_x))}{\partial_yg(k_x,k_y(k_x))}\,.
\label{delta-4}
\ee
This is an identity in virtue of (\ref{delta-2}) if $k_y$ takes the place of $x$.

Then we just proved formula (\ref{delta_app}) that we used during our calculation of the Jacobian of the change of variables evaluated in the FS.

\section*{ACKNOWLEDGMENTS: }
We would like to thank  G. Rossini for helpful discussions. This
work was partially supported by the ESF grant INSTANS, PICT ANCYPT (Grant No 20350), and PIP CONICET
(Grant No 5037).

%
%
%

\end{document}